\documentclass{ws-procs975x65}
\newcommand{\be}{\begin{equation}}
\newcommand{\ee}{\end{equation}}
\newcommand{\bea}{\begin{eqnarray}}
\newcommand{\eea}{\end{eqnarray}}
\newcommand{\lb}{\label}

\begin{document}



\title{SCALAR-TENSOR DARK ENERGY MODELS}

\author{R. GANNOUJI, D. POLARSKI, A. RANQUET}

\address{Lab. de Physique Th\'eorique et Astroparticules, CNRS\\
Universit\'e Montpellier II, 34095 Montpellier Cedex 05, France}

\author{A. A. STAROBINSKY}

\address{Landau Institute for Theoretical Physics, Moscow, 119334, Russia}


\begin{abstract}
We present here some recent results concerning scalar-tensor Dark Energy 
models. These models are very interesting in many respects: they allow for 
a consistent phantom phase, the growth of matter perturbations is modified. 
Using a systematic expansion of the theory at low redshifts, we relate the 
possibility to have phantom like DE to solar system constraints. 
\end{abstract}

\bodymatter

\section{Introduction}\label{intro}
The late-time accelerated expansion of the universe is a major challenge 
for cosmology. The component producing this acceleration accounts 
for about two thirds of the total energy density. While this has gradually 
become a building block of our present understanding, the nature of Dark 
Energy (DE) still remains mysterious \cite{SS00,Pad03,S05}. The simplest solution is 
a cosmological constant $\Lambda$. A major contender is Quintessence, a 
minimally coupled scalar field (with canonical kinetic term). 
We will consider scalar-tensor (ST) DE models, a more elaborate alternative 
involving a new physical degree of freedom, the scalar partner $\phi$ of the 
graviton responsible for a modification of gravity \cite{BEPS00,ST,EP01}. It is not 
clear yet whether some modification of gravity is required or even preferred 
in order to explain the bulk of data. The increasing accuracy of the data, 
should allow to severely constrain the various viable models. 
ST DE models allow for phantom DE, $w_{DE}<-1$, moreover the 
equation for the growth of matter perturbations is modified \cite{BEPS00}.  
We will review here results concerning their low $z$ behaviour, in particular 
how the DE equation of state is related to solar system constraints \cite{GPRS06}.  
\enlargethispage*{6pt}

\section{Scalar-tensor DE models}
We consider the microscopic Lagrangian density in the Jordan frame
\begin{equation}
L=\frac{1}{2} \Bigl (F(\Phi)~R -
Z(\Phi)~g^{\mu\nu}\partial_{\mu}\Phi\partial_{\nu}
\Phi \Bigr) - U(\Phi) + L_m(g_{\mu\nu})~.
\label{L}
\end{equation}
We {\it define} what we mean by the energy density 
$\rho_{DE}$ and the pressure $p_{DE}$ by writing the gravitational 
equations in the following Einsteinian form :
\bea
3F_0~H^2 &=& \rho_m + \rho_{DE} \lb{E1a} \\
-2F_0~{\dot H} &=&  \rho_m + \rho_{DE} + p_{DE} ~. \lb{E2a}
\eea
This can be seen as the {\em Einsteinian} form, with constant 
$G_0=G_N(t_0)= F_0^{-1}$, of the gravitational equations of ST gravity.
With these definitions, the usual conservation equation applies, 
and the equation of state parameter 
$w_{DE} \equiv \frac{p_{DE}}{\rho_{DE}}$ plays its usual role. 
Using (\ref{E1a},\ref{E2a}), 
One gets $w_{DE}(z)$ from the observations through
\be
w_{DE}(z) = \frac{\frac{1+z}{3} \frac{d h^2}{dz} - h^2 +\frac{1}{3}\Omega_{k,0}~(1+z)^2 }
                        {h^2 - \Omega_{m,0}~(1+z)^3 - \Omega_{k,0}~(1+z)^2 }~,\lb{wDE}
\ee
if we allow for a nonzero spatial curvature and $\Omega_m \equiv \frac{\rho_m}{3H^2 F_0}$. 

Looking at the equations above, everything looks the same as in 
GR, ST gravity is hidden in the definitions of $\rho_{DE}$, $p_{DE}$, 
and the various $\Omega$'s.  
The condition for DE to be of the phantom type, $w_{DE}<-1$, reads  
\be
\frac{d h^2}{dz} < 3 ~\Omega_{m,0}~(1+z)^2 + 2 ~\Omega_{k,0}~(1+z)~.\lb{ineq}
\ee
in the presence of spatial curvature \cite{SS00,BEPS00,PR05}. As first 
emphasized \cite{BEPS00}, the weak energy condition for DE can be 
violated in scalar-tensor gravity (see also \cite{T02}).

\section{General low $z$ expansion of the theory}
We investigate now the low $z$ behaviour of the model and the 
possibility to have phantom boundary crossing in a recent epoch. 
For each solution $H(z),~\Phi(z)$, the basic microscopic functions 
$F(\Phi)$ and $U(\Phi)$ can be expressed as functions of $z$ and 
expanded into
Taylor series in $z$:
\bea
\frac{F(z)}{F_0} = 1 + F_1 ~z + F_2 ~z^2 + ...> 0~,\lb{expz1}\\
\frac{U(z)}{3F_0~H_0^2}\equiv \Omega_{U,0}u = \Omega_{U,0} + u_1 ~z
+ u_2 ~z^2 + ...~.
\lb{expz2}
\eea
From (\ref{expz1},\ref{expz2}), all other expansions can be derived, in particular:
\bea
w_{DE}(z) &=& w_0 + w_1 ~z + w_2 ~z^2 + ...~,\\
H_0^{-1}\frac{ {\dot G}_{\rm eff} }{G_{\rm eff}} &=& g_0 + g_1~z +
g_2~z^2 + ....~.
\eea
A viable ST gravity model must be very close to General Relativity, viz.
\be
\omega_{BD,0} = \frac{6(\Omega_{DE,0} - \Omega_{U,0} - F_1)}{F_1^2}
= \frac{\Delta^2}{F_1^2} > 4 \times 10^4~, \lb{BD0}
\ee
with $\Delta^2\equiv 6~(\Omega_{DE,0}-\Omega_{U,0}-F_{1})$. 
Therefore, we must have $|F_1|\ll 1$ and $\Delta^2\approx
6(\Omega_{DE,0} - \Omega_{U,0})>0$.
Moreover, for positive $U$, $\Delta^2 < 6 \Omega_{DE,0}< 5$ 
\be
|F_1|<\left(\frac{5}{\omega_{BD,0}}\right)^{1/2}\lesssim 10^{-2}~.
\lb{F1}
\ee
It can be shown that the condition $|F_1|\ll 1$ is sufficient to ensure here
that solar system constraints are satified \cite{GPRS06}.  

We now specialize to the case $|F_1|\ll 1$ yet assuming that other
$F_i$ are not as small. Then all expansions simplify 
considerably and we have in particular,
\bea
1 + w_0 &\simeq&\frac{2F_{2} + 6(\Omega_{DE,0} - \Omega_{U,0})}
{3\Omega_{DE,0}}~. \lb{cp0a}
\eea
From (\ref{cp0a}), the necessary condition to have phantom DE {\it
today} reads
\be \left(\frac{d^2F}{d\Phi^2}\right)_0= \frac{F_2}{3~(\Omega_{DE,0} - \Omega_{U,0})} < -1 ~. \lb{cp2}
\ee
Hence $F_2<0$ is necessary for phantom DE, because $\Omega_{DE,0} - \Omega_{U,0}>0$ 
from $\Delta^2>0$. In addition significant phantom DE requires $|F_2|\sim 1$. 
If $|F_1|\sim |F_2|\ll 1$, the present phantomness is very small. 

It is actually possible to invert all expansions and to obtain all 
coefficients in function of the post-Newtonian parameters $\gamma,~\beta$ 
and $g_0$. The following results are finally obtained
\bea
F_1 &=& g_0~\frac{\gamma-1}{\gamma-1 - 4(\beta-1)}\\
F_2 &=& -2~g_0^2~ \frac{\beta-1}{[ \gamma-1 - 4(\beta-1) ]^2}\\
\Omega_{DE,0}-\Omega_{U,0} &=& -\frac{1}{6}~g_0^2~ \frac{\gamma-1}
                    {[ \gamma-1 - 4(\beta-1) ]^2}\\
1+w_{DE,0} &=& -\frac{1}{3}~g_0^2~\frac{4(\beta-1) + \gamma-1}{\Omega_{DE,0}~[ \gamma-1 - 4(\beta-1) ]^2}
\eea
The best present bounds are 
$\gamma_{PN}-1 = (2.1\pm 2.3)\cdot 10^{-5},~
\beta_{PN}-1 = (0\pm 1)\cdot 10^{-4},~
\frac{{\dot G}_{{\rm eff},0}}{G_{{\rm eff},0}} = (-0.2\pm 0.5)
\cdot 10^{-13}~y^{-1}$.
Though possible in principle \cite{MSU06}, the interesting possibility 
to test phantomness in the solar system is very hard while  
its amount depends critically on the small quantity $g_0^2$. 
In this respect cosmological data are certainly better suited, 
a conclusion reminiscent of that reached in \cite{EP01} concerning 
the viability of ST DE models with vanishing potential.

\vfill

\end{document}